\documentclass[preprint]{iucr}              

     \journalcode{J}

 \usepackage{mathtools}
  \usepackage{xcolor}
 \usepackage{algorithm}
 \usepackage{mymacros}
 \usepackage{dsfont}

\newlength{\hcolwidth}
\begin{document}

\title{General approaches for shear-correcting coordinate
  transformations in  Bragg coherent diffraction imaging: Part I}
\shorttitle{Shear-correcting coordinate  transformations in BCDI: Part I}
\date{}

\cauthor[a]{S.}{Maddali}{smaddali@anl.gov}{}
\author[b]{P.}{Li}
\author[c,d]{A.}{Pateras}
\author[e]{D.}{Timbie}
\author[a,f]{N.}{Delegan}
\author[e,g]{A. L.}{Crook}
\author[e]{H.}{Lee}
\author[a]{I.}{Calvo-Almazan}
\author[a]{D.}{Sheyfer}
\author[h]{W.}{Cha}
\author[a,e,f]{F. J.}{Heremans}
\author[a,e,f]{D. D.}{Awschalom}
\author[b]{V.}{Chamard}
\author[b]{M.}{Allain}
\author[a]{S. O.}{Hruszkewycz}

\aff[a]{Materials Science Division, Argonne National Laboratory, Lemont, IL 60439, USA}
\aff[b]{Aix-Marseille Univ, CNRS, Centrale Marseille, Institut Fresnel, Marseille, France}
\aff[c]{Center for Integrated Nanotechnologies, Los Alamos National Laboratory, Los Alamos, NM 87545, USA}
\aff[d]{Materials Science and Technology Division, Los Alamos National Laboratory, Los Alamos, NM 87545, USA}
\aff[e]{Pritzker School of Molecular Engineering, University of Chicago, 5640 S. Ellis Ave, Chicago, IL 60637, USA}
\aff[f]{Center for Molecular Engineering, Argonne National Laboratory, Lemont, IL 60439, USA}
\aff[g]{Dept. of Physics, University of Chicago, 5620 S. Ellis Ave, Chicago IL 60637}
\aff[h]{X-ray Science Division, Argonne National Laboratory, Lemont, IL 60439, USA}
\shortauthor{S. Maddali \textit{et al.}}
 
\begin{abstract}
	In this two-part article series we provide a generalized description of the scattering geometry of Bragg coherent diffraction imaging (BCDI) experiments, the shear distortion effects inherent to the resulting three-dimensional (3D) image currently used phase retrieval methods and strategies to mitigate this  distortion.
In this Part I, we derive in general terms the real-space coordinate transformation required to correct this shear, which has its origins in the more fundamental relationship between the mathematical representations of mutually conjugate 3D spaces.
Such a transformation, applied as a final post-processing step following phase retrieval, is crucial for arriving at an un-distorted, correctly oriented and physically meaningful image of the 3D crystalline scatterer. 
As the relevance of BCDI grows in the field of materials characterization, we take this opportunity to generalize the available sparse literature that addresses the geometric theory of BCDI and the subsequent analysis methods.
This geometrical aspect, specific to coherent Bragg diffraction and absent in two-dimensional transmission CDI experiments, gains particular importance when it comes to spatially-resolved characterization of 3D crystalline materials in a realiable, non-destructive manner.
This series of articles describes this theory, from the diffraction in Bragg geometry, to the corrections needed to obtain a properly rendered digital image of the 3D scatterer.
Part I of this manuscript provides the experimental BCDI community with the theoretical underpinnings of the 3D real-space distortions in the phase-retrieved object, along with the necessary post-retrieval correction method.
Part II builds upon the geometric theory developed in Part I with the formalism 
to correct the shear distortions directly on an orthogonal grid within the phase retrieval algorithm itself, allowing more physically realistic constraints to be applied.
Taken together, Parts I and II provide the x-ray science community with a set of generalized BCDI shear-correction techniques crucial to the final rendering of a 3D crystalline scatterer and for the development of new BCDI methods and experiments.

 \end{abstract}
\maketitle

\section{Introduction}
\label{S:intro}
Bragg coherent diffraction imaging (BCDI) is a lensless imaging method by which the morphology and internal strain state of compact crystalline objects may be visualized non-destructively \cite{Robinson2001,Robinson2009,Miao2015}. 
A 3D rendering of an appropriately oriented crystalline scatterer is obtained by coherently illuminating it with monochromatic x-rays, and computationally inverting the acquired 3D diffraction pattern using iterative phase retrieval algorithms \cite{Fienup1982,Fienup1987,Marchesini2003,Marchesini2007}.
The 3D coherent diffraction pattern is collected in the vicinity of a Bragg peak, using a pixelated area detector and by incrementally changing the preferred orientation of the object in the x-ray beam.
The method has recently seen increased use at third-generation synchrotron light sources for a variety of static, \emph{in situ} and \emph{operando} studies of materials in environments difficult to access with other characterization methods (\emph{e.g.} elevated temperatures or deeply embedded crystals)~\cite{Cha2016,Cherukara2018,Dupraz2015,Highland2017,Hofmann2017,Ulvestad2015,Ulvestad2015a}.

The goal of any coherent diffraction imaging measurement is to numerically compute the complex-valued field of the scattering object from the acquired diffraction.
Different variants of coherent diffraction imaging interpret this complex field in different ways. 
In particular, it represents the local complex refractive index in the case of two-dimensional (2D) transmission CDI experiments, and the local crystal lattice strain in the case of 3D Bragg scattering geometry.
In the specific case of BCDI (which includes beam-scanning techniques like Bragg ptychography~\cite{Hruszkewycz2017a}), a single component of the six-parameter strain tensor field within the scatterer bulk is encoded into the complex field of the retrieved object~\cite{Robinson2009}.
Unlike the refractive index which is a scalar invariant, lattice strain components depend intimately on the frame of reference.
Thus, accurate representation of the crystalline scatterer in a suitable real-space orthogonal frame is essential for the meaningful interpretation of the object morphology and strain state, and thereby the study of any physical process that may depend on these factors.

In BCDI, this orthogonal rendering is complicated by the fact that the diffraction signal, modeled as the squared modulus of the Fourier transform of the complex field of the scatterer~\cite{Goodman2005}, is necessarily sampled along non-orthogonal directions. 
This inevitably imposes a non-orthogonal shear on the axes of the 3D object array obtained from conventional phase retrieval, as we demonstrate in detail in Section~\ref{S:geom}.
The non-orthogonal sampling of the 3D space of the scattered signal (hereafter referred to as Fourier space) is intimately connected to experimental considerations such as (i) the pixel size of the area detector, (ii) object-detector distance, (iii) x-ray wavelength, (iv) the orientation of the mounted scatterer and (v) the manner of rotation (`rocking') of the scatterer in the x-ray beam.
Through the wave propagation process (here, the Fourier transform), the discrete sampling grid of the reconstructed real-space scatterer is also tightly constrained by these factors, and is non-orthogonal in general.
 A na\"ive 3D rendering of the phase retrieval result without accounting for this effect results in a sheared image not truly representative of the physical scatterer.
In this paper we rigorously derive the relationship between the two shears in real and Fourier space, and provide a prescription to correct the real-space distortion, enabling the correct 3D rendering of the scatterer.

Although BCDI has been steadily gaining popularity within the materials science community as a valuable nanoscale characterization method, existing literature on the underlying geometric theory is as yet sparse. 
Currently available literature consists of general-purpose tools to map Fourier space~\cite{Kriegner2013} and working-rule prescriptions for the post-processing of the phase retrieval output, tailored for  the highly specific experimental geometries of existing BCDI beamline facilities~\cite{Pfeifer2005,Pateras2015,Pateras2015a}.
In this paper, the first of two parts, we take the opportunity to fill this gap in the literature by providing an analysis of the intricate scattering geometry of a BCDI measurement, as well as deriving the general way to correct the 3D distortion. 
We achieve this by starting from basic considerations that analyze the conjugate nature of real and Fourier spaces themselves, and build up to the adaptation to discretely sampled fields, as in a real-world BCDI experiment. 

More specifically, Part I of this paper describes a method which takes as its input the geometric configuration of a BCDI experiment and returns a basis of three sampling vectors in 3D real space associated with the three independent axes of the phase retrieval solution array. 
This array (representing non-orthogonal samples of the scatterer), when combined with knowledge of the sampling basis, is sufficient to render a physically accurate (albeit shear-sampled) image of the scatterer with one of many available visualization tools such as Matlab, Python or Paraview.
Building on this foundation, Part II of this paper describes a formal derivation of a modification to the 3D Fourier transformation itself, appropriate for phase retrieval, in which the sheared Fourier-space sampling basis is directly used in the reconstruction of the scatterer on an orthogonal grid.

This paper is organized as follows: in Section~\ref{S:representation} we describe the mathematical representation of continuous 3D real and Fourier space points in generally non-orthogonal bases, and how these bases are related through the continuous Fourier transform.
Further, we adapt this conjugate relation between the bases to the case of the discrete Fourier transform, appropriate for BCDI experiments.
In Section~\ref{S:bcdi} we provide a generalized treatment of the scattering geometry generally applicable to any BCDI configuration. 
We then cast the physical quantities and transformation operators thus introduced into orthonormal coordinate frames convenient for realistic sample rendering, thereby demonstrating the ease of implementation of the theory developed in Section~\ref{S:representation} using standard software packages.
Section~\ref{S:example} describes a demonstrative example of a reconstructed 3D image of a silicon carbide (SiC) nano-particle from data collected at a BCDI facility, in which the computed shear correction is applied to the results of conventional phase retrieval. 
The image thus obtained is corroborated by SEM pictures of identically fabricated nano-particles.
In Section~\ref{S:summary} we close with a summary of our formalism and the shear-correction results.

\section{Representation of real and Fourier space}
\label{S:representation}
In a BCDI measurement, the discretization of three-dimensional Fourier space is achieved through (i) a pixelated area detector and (ii) finite angular steps in the `rocking' direction along which the scatterer is physically rotated. 
The sampling directions are determined by geometric considerations such as the placement of the detector (equivalently, the Bragg reflection of interest) and the direction of rocking of the scatterer.
These considerations in turn have a direct bearing on the subsequent discretization of the three-dimensional reconstructed object resulting from successful inversion from Fourier to real space.
This mirrors the more fundamental relation between continuous real and Fourier space representations and their connection through the continuous Fourier transform (CFT).
In this section we start by deriving this more fundamental relationship, and use it to arrive at the corresponding relation for the discrete case.

Before we begin, we describe the mathematical notation in use in this paper. 
Scalar quantities (real or complex) are denoted by lowercase, non-boldface Greek or Roman letters ($r$, $q$, $\psi$), 
while two- and three-dimensional vectors by lowercase boldface letters ($\boldsymbol{r}$, $\boldsymbol{q}$).
In addition, the Euclidean ($\ell^2$-) norm of a vector $\boldsymbol{r} $ is denoted by $\norm{\boldsymbol{r}}$ and vectors of unit norm are denoted by lowercase boldface letters with a caret ($\unitvector{s}_1$, $\unitvector{e}_1$, $\unitvector{k}_1$).
Matrices representing either rank-2 tensors (such as rotation operators) or three-dimensional basis sets are denoted by uppercase Roman letters, either in boldface ($\boldsymbol{B}$, $\boldsymbol{P}$) or in script font ($\mathcal{R}$, $\mathcal{D}$, $\mathcal{I}$). 
The determinant of a square matrix $\boldsymbol{B}$ is denoted  by $\det(\boldsymbol{B})$. 

A three-dimensional real-space vector $\boldsymbol{v}$ denoting position is represented 
in a given ``reference'' orthonormal frame as a linear combination of an ordered set of basis 
vectors $\left[ \unitvector{s}_1~ \unitvector{s}_2~ \unitvector{s}_3\right]$ of unit norm:
\begin{equation}
	\boldsymbol{v} := \sum\limits_{n=1}^3 r_n \unitvector{s}_n
	\label{eq.vector}
\end{equation}
In our convention, the coefficients $r_n \in \mathds{R}$ carry dimensions of 
length while  $\{\unitvector{s}_n\}_{n=1}^3$ merely denote a set of three 
orthogonal directions with unit norms. If the components 
of each basis vector $\unitvector{s}_n$ (expressed in the same frame) are 
concatenated as the columns of a $3 \times 3$ matrix $\boldsymbol{L}$, 
then Eq.~\eqref{eq.vector} can be written compactly:
\begin{equation}
	\boldsymbol{v} := \boldsymbol{L} \boldsymbol{r}
	\label{eq.vector_linop}
\end{equation}
where $\boldsymbol{r} \equiv \left[r_1~r_2~r_3\right]^T$. 
Here `$T$' denotes the matrix transpose.
The same vector $\boldsymbol{v}$ 
can be expressed in another basis $\tilde{\boldsymbol{L}}$ of non-coplanar unit vectors 
$[\hat{\tilde{\boldsymbol{s}}}_1 ~ \hat{\tilde{\boldsymbol{s}}}_2 ~\hat{\tilde{\boldsymbol{s}}}_3]$ with a 
different set of coefficients $\tilde{\boldsymbol{r}} = \left[\tilde{r}_1~\tilde{r}_2~\tilde{r}_3\right]^T$:
\begin{equation}
	\boldsymbol{v} = \tilde{\boldsymbol{L}} \tilde{\boldsymbol{r}}
	\label{eq.vector_linop2}
\end{equation}
From eqs.~\eqref{eq.vector_linop} and~\eqref{eq.vector_linop2}, the prescription for transformation between two representations of the vector $\boldsymbol{v}$ can be derived:
\begin{equation}
	\boldsymbol{r} = \boldsymbol{B}_r \tilde{\boldsymbol{r}}
	\label{eq.realtransform}
\end{equation}
where $\boldsymbol{B}_r \equiv {\boldsymbol{L}^{-1}\tilde{\boldsymbol{L}}}$ denotes the linear transformation operator from the ``tilded'' to the ``un-tilded'' basis.
We now consider a scalar field in real space, which is represented by two distinct but equivalent scalar functions $\psi: \mathds{R}^3 \rightarrow \mathds{C}$ and $\tilde{\psi}: \mathds{R}^3 \rightarrow \mathds{C}$ 
such that 
\begin{equation}
\psi(\boldsymbol{r}) = \tilde{\psi}(\tilde{\boldsymbol{r}}). 
\label{equiv_realspace}
\end{equation}
Let the Fourier transforms of the functions $\psi(\boldsymbol{r})$ and $\tilde{\psi}(\tilde{\boldsymbol{r}})$ respectively be $\Psi(\boldsymbol{q})$ and $\tilde{\Psi}(\tilde{\boldsymbol{q}})$.
These Fourier transforms have a specific interpretation in the BCDI problem considered in the next section. 
Specifically, $\tilde\Psi$ is the distorted representation of the scattered 3D wave field whose intensity distribution is probed during the BCDI measurement whereas $\Psi$ is the undistorted representation of the field in its orthogonal frame.
Here, $(\boldsymbol{r}, \boldsymbol{q})$ and $(\tilde{\boldsymbol{r}}, \tilde{\boldsymbol{q}})$ are conjugate pairs of vector coordinates as defined by the Fourier transform.
Then for $\Psi(\boldsymbol{q})$ and $\tilde{\Psi}(\tilde{\boldsymbol{q}})$ to represent the same physical scalar field in Fourier space, 
the quantities $\boldsymbol{q}$ and $\tilde{\boldsymbol{q}}$ should satisfy the following conditions: 
\begin{enumerate}
	\item	They should be two distinct three-component representations of the same physical Fourier-space point.
	\item They should be related by a linear transformation akin to Eq.~\eqref{eq.realtransform}:
          \begin{equation}
            \label{eq.reciptransform}
            \boldsymbol{q} \equiv \boldsymbol{B}_q \tilde{\boldsymbol{q}}.
          \end{equation}
\end{enumerate}
As with real space, we adopt the convention that the elements of the column matrix representation $\boldsymbol{q}$ carry physical units of inverse length and the columns of $\boldsymbol{B}_q$ are dimensionless directions in Fourier space.
We note that in this paper we favor the `strict reciprocal' convention ($q \sim 1 / r$) over the more familiar solid state physicists' convention that carries an additional multiplicative constant ($q \sim 2\pi / r$). 
The convention that we adopt brings the real- and Fourier-space coordinates on an equal footing. 
Their phase relation is explicitly written as a scaling of $2\pi$ in the expression for the symmetric Fourier transform that we use hereafter in this paper.
The linear operator $\boldsymbol{B}_q$ is determined in terms of $\boldsymbol{B}_r$ in the following manner:
\begin{align}
  \tilde{\Psi}(\tilde{\boldsymbol{q}})  
  &:= \int_{\mathds{R}^3} \tilde{\psi}(\tilde{\boldsymbol{r}})\, e^{-\iota 2\pi {\tilde{\boldsymbol{r}}}^T \tilde{\boldsymbol{q}}}\, \mathrm{d}\tilde{\boldsymbol{r}}\notag \\
  & = \int_{\mathds{R}^3}  \tilde{\psi}(\tilde{\boldsymbol{r}}) \,e^{-\iota 2\pi (\boldsymbol{B}_r \tilde{\boldsymbol{r}})^T\boldsymbol{B}_r^{-T} {\tilde{\boldsymbol{q}}}} \, \mathrm{d} \tilde{\boldsymbol{r}}  \notag 
\end{align}
with $\iota:=\sqrt{-1}$ and the `$-T$' superscript denoting the inverse of the transpose, or equivalently the transpose of the inverse. 
By changing the integration variable from $\tilde{\boldsymbol{r}}$ to $\boldsymbol{r}$ according to Eq.~\eqref{eq.realtransform} we obtain
\begin{align}
  \tilde{\Psi}(\tilde{\boldsymbol{q}})
  &= \frac{1}{\det(\boldsymbol{B}_r)} \int_{\mathds{R}^3}  \tilde{\psi}(\boldsymbol{B}^{-1}_r \boldsymbol{r}) \,e^{-\iota 2\pi \boldsymbol{r}^T\boldsymbol{B}_r^{-T} \tilde{\boldsymbol{q}}} \, \mathrm{d} {\boldsymbol{r}}  \notag \\
  &=  \frac{1}{\det(\boldsymbol{B}_r)}  \int_{\mathds{R}^3}  \psi (\boldsymbol{r}) \,e^{-\iota 2\pi {\boldsymbol{r}}^T\boldsymbol{B}_r^{-T} \tilde{\boldsymbol{q}}} \, \mathrm{d} {\boldsymbol{r}}  \notag \\
  &=  \frac{1}{\det(\boldsymbol{B}_r)}   \Psi(\boldsymbol{B}_r^{-T} \tilde{\boldsymbol{q}}) 
                                         \label{eq.equivalence}
\end{align}
where the relation $\tilde{\psi}(\boldsymbol{B}^{-1}_r \boldsymbol{r}) = \psi(\boldsymbol{r})$ 
(derived from Eqs.~\eqref{eq.realtransform} and \eqref{equiv_realspace}) was used in the second equality.
A direct consequence of \eqref{eq.equivalence} is that $\boldsymbol{q}$ and $\boldsymbol{B}_r^{-T} \tilde{\boldsymbol{q}}$ 
are actually the same physical point in the Fourier space: as a result, we have $\boldsymbol{q} = \boldsymbol{B}_r^{-T} 
\tilde{\boldsymbol{q}}$ from which we deduce with Eq.~\eqref{eq.reciptransform}    
\begin{equation}
	 \boldsymbol{B}_q = \boldsymbol{B}_r^{-T}.
	\label{eq.continuoustransform}
\end{equation}
By definition, representations of mutually conjugate spaces in a Fourier sense obey Eq.~\eqref{eq.continuoustransform}.
This relation was first alluded to in the Ph.D thesis of Anastasios Pateras~\cite{Pateras2015}.
It tells us that if the pair of variables $\left(\tilde{\boldsymbol{r}}, \tilde{\boldsymbol{q}}\right)$ are 
Fourier-conjugate to each other, then so are $(\boldsymbol{B}_r\tilde{\boldsymbol{r}}, \boldsymbol{B}_r^{-T}\tilde{\boldsymbol{q}})$, 
or equivalently, $(\boldsymbol{B}_q^{-T}\tilde{\boldsymbol{r}}, \boldsymbol{B}_q\tilde{\boldsymbol{q}})$.
Furthermore, if $\boldsymbol{B}_r$ represents an orthonormal frame, $\boldsymbol{B}_r$ is an orthogonal matrix (\textit{i.e.,} $\boldsymbol{B}_r^{-1} = \boldsymbol{B}_r^T$) and \eqref{eq.continuoustransform} reads $\boldsymbol{B}_r = \boldsymbol{B}_q$, \emph{i.e.} orthonormal bases defined in this manner are self-congujate. 

Consider the case applicable to BCDI measurements in which Fourier space is sampled in integer multiples of step sizes $\left(\delta q_1, \delta q_2, \delta q_3\right)$ along directions specified by the columns of a basis matrix $\boldsymbol{B}_q$. Each position in Fourier space is indexed by the integer vector $\boldsymbol{n} \equiv \left[i~j~k\right]^T$ such that:
\begin{equation}
	\boldsymbol{q} = \boldsymbol{B}_q 
	\underbrace{
		\left[
			\begin{matrix}
				\delta q_1 &&\\&\delta q_2&\\&&\delta q_3
			\end{matrix}
		\right]
	}_{\boldsymbol{\Lambda}_q} 
	\left[
		\begin{matrix}
			i\\j\\k
		\end{matrix}
	\right] = \boldsymbol{B}_q \boldsymbol{\Lambda}_q \boldsymbol{n}. 
	\label{eq.recip_discrete}
\end{equation}
Here $\boldsymbol{\Lambda}_q$ is a diagonal matrix whose elements carry physical dimensions of inverse length.
For convenience, we write the discretization as $\boldsymbol{q} = \boldsymbol{B}_\text{recip} \boldsymbol{n} $ in terms of $\boldsymbol{n} $ and the individual Fourier-space steps determined by the columns of the matrix $\boldsymbol{B}_\text{recip} := \boldsymbol{B}_q \boldsymbol{\Lambda}_q$.
Similarly, the discretization of real space is parameterized by an integer vector $\boldsymbol{m}  \equiv \left[l~m~n\right]^T$ as:
\begin{equation}
	\boldsymbol{r} = \boldsymbol{B}_r
	\underbrace{
		\left[
			\begin{matrix}
				\delta r_1 &&\\&\delta r_2&\\&&\delta r_3
			\end{matrix}
		\right]
	}_{\boldsymbol{\Lambda}_r} 
	\left[
		\begin{matrix}
			l\\m\\n
		\end{matrix}
	\right] = \boldsymbol{B}_r \boldsymbol{\Lambda}_r \boldsymbol{m}.
	\label{eq.real_discrete}
\end{equation}
Here we similarly define $\boldsymbol{B}_\text{real} := \boldsymbol{B}_r \boldsymbol{\Lambda}_r$ whose columns denote individual real-space steps.
We now wish to approximate the CFT using the discrete Fourier transform (DFT) on a grid of size $N_1 \times N_2 \times N_3$. 
Provided the fringe intensity distribution is sufficiently sampled and the experimental Fourier-space aperture is large enough to avoid cyclic aliasing issues, the following approximation is true of the complex phase factor $\boldsymbol{q}^T \boldsymbol{r}$:
\begin{align}
  \boldsymbol{q}^T \boldsymbol{r} 
  &= 
	\left(\frac{il}{N_1} + \frac{jm}{N_2} + \frac{kn}{N_3}\right) \label{eq.phaserelation} \\
	\Longrightarrow \left(\boldsymbol{n}^T \boldsymbol{B}_\text{recip}^T \right)\left(\boldsymbol{B}_\text{real} \boldsymbol{m} \right)
  &= \boldsymbol{n}^T \underbrace{\left[\begin{matrix}N_1^{-1}&&\\&N_2^{-1}&\\&&N_3^{-1}\end{matrix}\right]}_{\mathcal{D}} \boldsymbol{m} \tag{$\forall \boldsymbol{m}, \boldsymbol{n} \in \mathds{Z}^3$} \\
		\Longrightarrow \boldsymbol{B}_\text{recip}^{-T} \mathcal{D} \boldsymbol{B}_\text{real}^{-1} 
	&= \mathcal{I}
		\label{eq.discretetransform}
\end{align}
where $\mathcal{I}$ is the $3 \times 3$ identity matrix.
The identity in Eq.~\eqref{eq.discretetransform} follows from the fact that Eq.~\eqref{eq.phaserelation} holds for \emph{all} integer vectors $\boldsymbol{n},~\boldsymbol{m} \in \mathds{Z}^3$.
Eq.~\eqref{eq.discretetransform} is the discrete analog of Eq.~\eqref{eq.continuoustransform}. 
At this point we note its equivalence to other documented prescriptions that relate conjugate sampling bases of the general form 
\begin{equation}
\left\{
\begin{matrix}
	\boldsymbol{B}_\text{real} &\equiv& [\unitvector{e}_1^{\prime}~~\unitvector{e}_2^{\prime}~~ \unitvector{e}_3^{\prime}] \boldsymbol{\Lambda}_r \\
	\boldsymbol{B}_\text{recip} &\equiv& [\unitvector{k}_1^{\prime}~~\unitvector{k}_2^{\prime}~~\unitvector{k}_3^{\prime}] \boldsymbol{\Lambda}_q.
\end{matrix}
\right. 
\end{equation}
Here the primed quantities $\unitvector{e}_i^\prime$ and $\unitvector{k}_i^\prime$ denote unit-norm sampling directions (not necessarily mutually orthogonal) in real and Fourier space respectively.
The columns of $\boldsymbol{B}_\text{real} = \boldsymbol{B}_\text{recip}^{-T} \mathcal{D}$ are indeed given by:
\begin{align}
	\delta r_1 \unitvector{e}_1^{\prime} &=  \frac{1}{N_1}\frac{\delta q_2 \unitvector{k}_2^{\prime} \times \delta q_3 \unitvector{k}_3^{\prime}}{\left(\delta q_1 \delta q_2 \delta q_3\right)\unitvector{k}_1^{\prime} \cdot \unitvector{k}_2^{\prime} \times \unitvector{k}_3^{\prime}} = \frac{1}{V_{123}} N_2 \delta q_2 \unitvector{k}_2^{\prime} \times N_3 \delta q_3 \unitvector{k}_3^{\prime} \label{eq.x1} \\
	\delta r_2 \unitvector{e}_2^{\prime} &= \frac{1}{V_{123}} N_3 \delta q_3 \unitvector{k}_3^{\prime} \times N_1 \delta q_1 \unitvector{k}_1^{\prime} \label{eq.x2} \\
	\delta r_3 \unitvector{e}_3^{\prime} &= \frac{1}{V_{123}} N_1 \delta q_1 \unitvector{k}_1^{\prime} \times N_2 \delta q_2 \unitvector{k}_2^{\prime} \label{eq.x3}
\end{align}
where `$\cdot$' and `$\times$' denote the dot-product and cross-product respectively,  and $V_{123} : = \left(N_1\delta q_1\right)\left( N_2\delta q_2\right)\left( N_3\delta q_3\right) \unitvector{k}_1^{\prime} \cdot \unitvector{k}_2^{\prime} \times \unitvector{k}_3^{\prime} = \det\left(\boldsymbol{B}_\text{recip} \mathcal{D}^{-1}\right)$ is the total Fourier-space volume queried over the entire BCDI scan.
Up to the conventional multiplicative factor of $2\pi$ mentioned earlier, Eqs.~\eqref{eq.x1}, \eqref{eq.x2} and \eqref{eq.x3} are identical to the several familiar prescriptions for coordinate inversions found in the existing literature~\cite{Pfeifer2005,Berenguer2013,Pateras2015,Yang2019}.

In addition, they are reminiscent of the conversion between the primitive vectors of an atomic crystal's real- and reciprocal-space Bravais lattices, from solid-state physics.
Up to the effect of the finite Fourier-space volume (represented by the scaling factor of $1/N_i$ in each dimension), the relationship between $\boldsymbol{B}_\text{real}$ and $\boldsymbol{B}_\text{recip}$ mirrors that between the primitive vectors of these Bravais lattices.
This is because both relationships have their origins in the underlying concept of far-field coherent diffraction from an array of regularly spaced point scatterers.
In the context of BCDI phase retrieval, these point scatterers represent digitized samples of a numerical diffracting object, while in a physical lattice they represent actual atomic electron clouds.
\vspace{.5em}

In a typical BCDI geometry, additional complications are introduced owing to the fact that the location of a Bragg reflection of interest (and therefore the accompanying coherent diffraction pattern), is in general offset with respect to the absolute origin of a universal frame. Such offsets result in phase effects in the corresponding real-space reconstructed object. 
We briefly address this issue of additional phase contributions and show that they in fact have no effect on a measured BCDI signal.
Ref.~\cite{Vartanyants2001} contains an \emph{ab initio} treatment of the various phase effects in a BCDI diffracted wave field as a result of origin offsets.  
We consider a discrete sample point $\boldsymbol{q}$ in Fourier space in relation to an (as 
yet unspecified) origin  in the vicinity of a Bragg peak located at $\boldsymbol{q}_0$:
\begin{equation}\label{eq.qstep}
	\boldsymbol{q} = \boldsymbol{q}_0 + \boldsymbol{B}_\text{recip} \boldsymbol{n}.
\end{equation}
Analogously in real space, the complex-valued scatterer is sampled at points $\boldsymbol{r}$ given by:
\begin{equation}\label{eq.xstep}
	\boldsymbol{r} = \boldsymbol{r}_0 + \boldsymbol{B}_\text{real} \boldsymbol{m}.
\end{equation}
In this formulation, $\boldsymbol{r}_0$ and $\boldsymbol{q}_0$ are chosen as arbitrary constant offsets in real and Fourier space, even though $\boldsymbol{q}_0$ is in fact determined by the Bragg scattering geometry, as we shall see in Sec.~\ref{S:bcdi}.
The complex phase factor now becomes:
\begin{equation}
	\boldsymbol{q}^T \boldsymbol{r} = 
		\boldsymbol{q}_0^T \boldsymbol{r}_0 + 
		\boldsymbol{q}_0^T \boldsymbol{B}_\text{real}\boldsymbol{m} + 
		\boldsymbol{n}^T\boldsymbol{B}_\text{recip}^T \boldsymbol{r}_0 + 
		\boldsymbol{n}^T\boldsymbol{B}_\text{recip}^T \boldsymbol{B}_\text{real}\boldsymbol{m}.
\end{equation}
Here each term may be qualitatively understood as follows:
\begin{enumerate}
	\item	The first term is a constant phase term and has no effect on the measured intensity: $I \propto \left|\Psi\right|^2$
	\item	The second term introduces a constant offset in the absolute position of the measured diffraction pattern but does not change the measured intensity distribution. In practice, this term is set to zero by enforcing that the maximum of the Bragg peak is centered in the numerical array.
	\item	The third term introduces a phase ramp in $\Psi$ that encodes the translation of the scatterer, but it does not affect the measured intensity distribution.
	\item	The fourth term results from the discrete sampling of real and Fourier spaces.
\end{enumerate}

We therefore see that as far as the measured intensity distribution is concerned, the constant real- and Fourier-space offsets $\boldsymbol{q}_0$ and $\boldsymbol{r}_0$ characteristic of a BCDI measurement may be set to zero without loss of generality.
This allows us to apply Eq.~\eqref{eq.discretetransform} directly to the BCDI sampling bases in real and Fourier space.

We note that the method developed to compute $\boldsymbol{B}_\text{real}$ for use in Eq.~\eqref{eq.xstep} merely seeks to associate a sheared sampling basis with the three independent axes of the phase retrieval solution array.
The actual rendering of the physically accurate scatterer (albeit on a sheared sampling grid) may be achieved with one of many available software packages for 3D visualization.
One potential shortcoming of this rendering convention is the subsequent computation of local lattice strain in the crystalline scatterer, which requires evaluating the spatial gradient of the complex phase at these sheared grid points, in non-rectilinear coordinates.
Under these circumstances, the complex phase in real space may first be approximated at the nodes of a new orthogonal grid \emph{via} interpolation, followed by the usual computation of the gradient in rectilinear coordinates~\cite{Newton2009,Hofmann2017a}.
Alternatively, one may do away with real-space interpolation altogether and directly compute the correct strain component at each non-rectilinear grid point.
For the interested reader, we derive this latter computation in Appendix~\ref{S:strain}.

\section{Quantitative aspects of BCDI}
	\label{S:bcdi}

Having established the use of Eq.~\eqref{eq.discretetransform} for the purposes of BCDI, we proceed to a general description of the geometry of a BCDI measurement. 
In Section~\ref{S:prelim} we first provide a symbolic, frame-agnostic description of the relevant degrees of freedom and vector quantities a BCDI experiment.
In Section~\ref{S:coords} we describe the relevant orthonormal coordinate frames in which to analytically represent these quantities and cast the subsequent discussion on BCDI geometry that is the subject of this article.
In Section~\ref{S:geom} we finally derive the analytical expressions of the relevant vector quantities and rotation operators, with respect to the appropriate coordinate frame.
We refer to the schematic in Fig.~\ref{fig:coord_frame}. 
	\subsection{Scattering preliminaries}
	\label{S:prelim}
	3D BCDI data sets are obtained by illuminating an isolated single-crystal scatterer with a coherent x-ray beam and rotating it about a fixed axis in small steps. 
The face of the detector is typically aligned perpendicular to the exit beam and defines the Fourier-space measurement plane.
Each sample rotation increment slightly displaces the measurement plane relative to its previous position and relative to the center of the Bragg reflection itself.
In this manner, the diffraction pattern is measured slice by slice in Fourier space, resulting in a 3D data array with indices $\boldsymbol{n}= [i~ j ~ k]$, where $i$ and $j$ correspond to the pixel coordinates of the detector, and $k$ corresponds to angular increments. 
A typical size for this data array is $\sim 256 \times 256 \times 64$~\cite{Cha2016}.

In any given 2D detector image from such a BCDI data set, oversampling of the fringe intensity pattern is achieved at hard x-ray wavelengths ($\lambda \simeq 0.1$ nm) by using a fine-pixel-pitch detector positioned $\sim 1$ m from the sample. 
In the third direction, fringe oversampling is enforced through sufficiently small rotational increments of the scatterer ($\sim 0.01^\circ$).  
Because the angular step is sufficiently small and the area detector subtends a very small portion of the Ewald sphere, the measurement planes can be considered parallel in Fourier space, as depicted in Fig.~\ref{fig:concept}(b). 
Though parallel, the measured slices are not sampled in an orthogonal manner, as we shall see with the explicit derivation of the sampling vectors $\boldsymbol{q}_i$, $\boldsymbol{q}_j$ and $\boldsymbol{q}_k$. 

An arbitrary point in Fourier space is determined on an absolute scale by $\boldsymbol{q} = \boldsymbol{k}_f - \boldsymbol{k}_i$ where $\boldsymbol{k}_i$ and $\boldsymbol{k}_f$ are the wave vectors of the incident and scattered x-rays respectively and $\norm{\boldsymbol{k}_i} = \norm{\boldsymbol{k}_f} = 1/\lambda$, the reciprocal of the x-ray wavelength.
One such point $\boldsymbol{q}_0$ corresponds to the center of the Bragg reflection, a location easily identified in BCDI data as the peak of the intensity distribution.
$\boldsymbol{q}_0$ sweeps through a small angle $\Delta\Omega$ between successive image acquisitions.
In the Fig.~\ref{fig:coord_frame} schematic, the crystal is rotated about the $\unitvector{s}_2$ direction, as is common practice at conventional BCDI facilities like the 34-ID-C end station of the Advanced Photon Source.
The resulting displacement of the measurement plane with respect to the diffraction pattern has a magnitude $\norm{\boldsymbol{q}_0} \Delta\Omega$ in Fourier space.
This quantity, equal to $\norm{\boldsymbol{q}_k}$ (\emph{i.e.} the third Fourier space sampling vector)  is derived explicitly in Section~\ref{S:geom}.

The discretized sampling of the relative Fourier-space position $\boldsymbol{q} - \boldsymbol{q}_0$ as a result of the pixel measurements and the rotational positions of the scatterer can be written in a consolidated manner: $\boldsymbol{q} - \boldsymbol{q}_0 = \boldsymbol{B}_\text{recip} \boldsymbol{n}$, where $\boldsymbol{B}_\text{recip} = \left[\boldsymbol{q}_i~\boldsymbol{q}_j~\boldsymbol{q}_k\right]$ comes from Eq.~\eqref{eq.qstep}. 
Further, we note two characteristics that hold for BCDI measurements:
\begin{enumerate}
	\item	
		When the detector face is oriented along the measurement plane, we have $\boldsymbol{q}_i \perp \boldsymbol{q}_j$ but the Bragg scattering geometry ensures that both are never simultaneously perpendicular to $\boldsymbol{q}_k$. 
		This is proved rigorously in Section~\ref{S:geom} and is the reason for the sheared sense of Fourier-space sampling.
	\item
		The norms of these sampling vectors in Fourier space are given by: 
		\begin{align}
			\norm{\boldsymbol{q}_i} &= \norm{\boldsymbol{q}_j} = p / \lambda D \label{eq.qiqj} \\
			\norm{\boldsymbol{q}_k} &= \norm{\boldsymbol{q}_0}\Delta\Omega = 2(\Delta \Omega) \sin\theta_B / \lambda \label{eq.dOmega}
		\end{align}
		where $p$ is the physical pixel size, $\lambda$ is the wavelength of illumination, $\theta_B$ is the Bragg angle of scattering, $D$ is the object-detector distance and $\Delta\Omega$ is the magnitude of the angle swept by $\boldsymbol{q}_0$ due to the rotation of the crystal by a single angular increment. 
		The numerical value of $\Delta\Omega$ is specific to a given diffractometer setup.
\end{enumerate}

	\subsection{Coordinate conventions}
	\label{S:coords}
	The vector and matrix quantities introduced thus far in Section~\ref{S:prelim} are symbolic in nature without explicit representation in a coordinate frame, and the relations between them are true for any BCDI configuration.
We now enumerate the bases in which these quantities are most naturally expressed in order to develop the numerical machinery for our demonstrative examples.
The frames we define are seen in Fig.~\ref{fig:coord_frame}:
\begin{enumerate}
	\item	
		We choose as a reference frame the synchrotron-based orthonormal laboratory frame denoted by the matrix of column vectors of unit norm: $\boldsymbol{B}_\text{lab} \equiv \left[\unitvector{s}_1~\unitvector{s}_2 ~\unitvector{s}_3\right]$ in which $\unitvector{s}_3$ points along the incident beam (downstream) and $\unitvector{s}_2$ points vertically upward. This is the orthonormal frame chosen for the display of the final BCDI reconstruction.
	\item
		A second frame $\boldsymbol{B}_\text{det} \equiv [\unitvector{k}_1~\unitvector{k}_2~\unitvector{k}_3]$ is attached to the detector.
		This frame is instrumental in determining the first two of the three sampling vectors $(\boldsymbol{q}_i, ~\boldsymbol{q}_j, ~\boldsymbol{q}_k)$, as we shall demonstrate presently.
		Two of the three mutually orthogonal directions of this frame lie in the measurement plane, while the third one is perpendicular to it, in the direction of the (nominal) exit beam.
		When the detector face is aligned with the measurement plane (\emph{i.e.} the detector is perpendicular to the exit beam), the directions of $\boldsymbol{q}_i$ and $\boldsymbol{q}_j$ coincide with the axes of this frame.
\end{enumerate}

		This second coordinate frame is seen to be used in several works in BCDI and Bragg ptychography~\cite{Cha2016,Hruszkewycz2012,Hruszkewycz2017a,Hruszkewycz2017}.
		In transmission mode (\emph{i.e.} the direct beam is incident upon the detector), $\boldsymbol{B}_\text{det}$ coincides exactly with $\boldsymbol{B}_\text{lab}$ in terms of orientation.
		At the 34-ID-C end station of the Advanced Photon Source (dedicated to BCDI measurements), the detector placement from transmission mode to Bragg mode is achieved with two rotational motors. 
		This corresponds to a two-parameter transformation (\emph{i.e.} corresponding to the $\gamma$ and $\delta$ angular rotations from Fig.~\ref{fig:coord_frame}) that takes the axes of the frame $\boldsymbol{B}_\text{lab}$ to the position $\boldsymbol{B}_\text{det}$. 
		We derive the general expression for this transformation in Section~\ref{S:geom}.

Central to this example and indeed to the BCDI geometry in general is the numerical representation of matrix operators denoting active rotations. 
We now provide a known prescription to compute such a $3 \times 3$ rotation matrix from knowledge of the angle of rotation and the direction about which the rotation is taking place (the axis-angle representation).
Such matrices are used extensively in the next section and are a convenient aid to computing rotation operators for any BCDI configuration.

We consider the active rotation of a vector $\boldsymbol{v}$ (expressed in some convenient frame such as the laboratory frame $\boldsymbol{B}_\text{lab}$) by an angle $\alpha$, about a unit-norm axis $\unitvector{u} \equiv \left[u_1~u_2~u_3\right]^T\in\mathds{R}^3$, \emph{in a right-handed or counterclockwise sense}. 
Here $\unitvector{u}$ is expressed in the same orthogonal frame as $\boldsymbol{v}$. 
The scalar $\alpha$ is invariant in different frames, while $\norm{\unitvector{u}} = 1$.
The rotation matrix is then given by~\cite{Rodrigues1840}:
\begin{align}
	 \mathcal{R}(\alpha, \unitvector{u}) &= 
		(\cos\alpha) \mathcal{I} + 
		(1 - \cos\alpha)\unitvector{u}\unitvector{u}^T + 
		(\sin\alpha) \boldsymbol{S}_{\unitvector{u}}
	\label{eq.rotmatrix} 
\end{align}
where $\unitvector{u} \unitvector{u}^T$  is the projector onto  $\unitvector{u}$ and
\begin{equation}
  \boldsymbol{S}_{\unitvector{u}} 
  = \left[
    \begin{matrix}
      0 & -u_3 & u_2 	\\ 
      u_3 & 0 & -u_1 	\\
      -u_2 & u_1 & 0
    \end{matrix}
  \right]
  \nonumber
\end{equation}
is the skew-symmetric matrix constructed from components of $\unitvector{u}$, or equivalently the operator version of the cross-product: $\boldsymbol{S}_{\unitvector{u}} \boldsymbol{v} = \unitvector{u} \times \boldsymbol{v}~~\forall \boldsymbol{v} \in \mathds{R}^3$.
Eq.~\eqref{eq.rotmatrix} is used frequently in our numerical examples in Section~\ref{S:geom}, and we provide it here as an aid to compute rotation matrices for a variety of different scattering and rocking geometries.
In the following analysis, we take the notation `$\mathcal{R}(\alpha, \unitvector{u}) \boldsymbol{v}$' to denote the resultant vector when the rotation matrix $\mathcal{R}(\alpha, \unitvector{u})$ acts on the column vector $\boldsymbol{v}$, with the understanding that the components of both $\unitvector{u}$ and $\boldsymbol{v}$ are expressed in the same frame.
 	
	\subsection{Sampling geometry}
	\label{S:geom}
\begin{figure}
	\centering
    \includegraphics[width=0.65\textwidth]{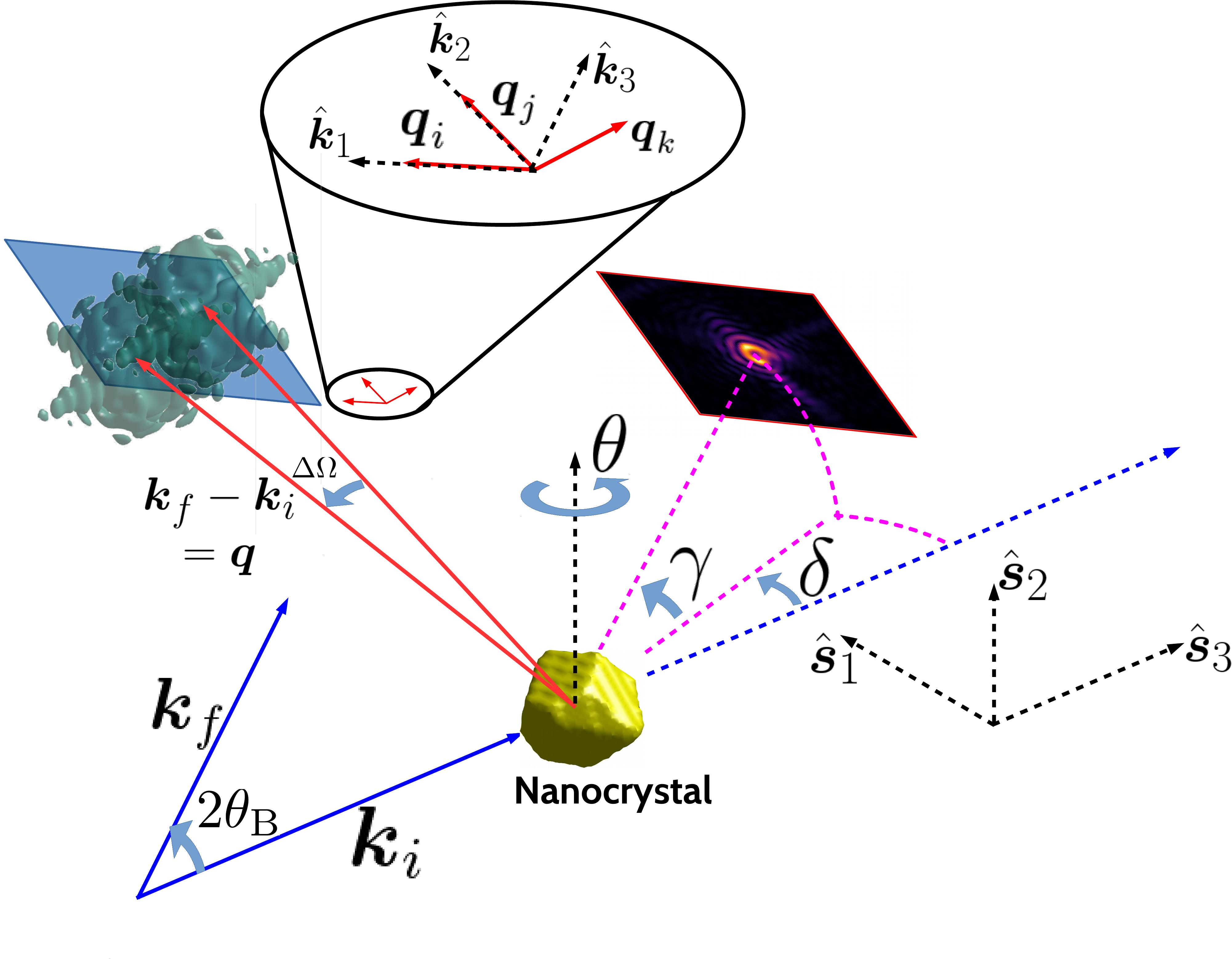}
    \caption{
    	Basic geometry of a BCDI measurement.
        Also shown are the laboratory frame $\boldsymbol{B}_\text{lab} \equiv [\unitvector{s}_1~\unitvector{s}_2~\unitvector{s}_3]$, the detector frame $\boldsymbol{B}_\text{det} \equiv [\unitvector{k}_1~\unitvector{k}_2~\unitvector{k}_3]$ and the sampling basis for Fourier space imposed by the scattering and object rotation geometry: $\boldsymbol{B}_\text{recip} \equiv [\boldsymbol{q}_i~\boldsymbol{q}_j~\boldsymbol{q}_k]$. 
        The $\gamma$ and $\delta$ degrees of freedom are specific to the 34-ID-C end station of the Advanced Photon Source.
    }
   	\label{fig:coord_frame}
\end{figure}

	In the laboratory frame, the axes of the laboratory frame itself are trivially expressed as the columns of the identity matrix:
\begin{equation}
	\boldsymbol{B}_\text{lab} = \mathcal{I}
\end{equation}
Put another way, in the laboratory frame, $\unitvector{s}_1 = [1~0~0]^T$, $\unitvector{s}_2 = [0~1~0]^T$ and $\unitvector{s}_3 = [0~0~1]^T$.
Then from Fig.~\ref{fig:coord_frame} (the arrangement at 34-ID-C), the orientation of the detector frame $\boldsymbol{B}_\text{det}$ is achieved by an active rotation of the laboratory frame, which is composed of two rotations of the type denoted in Eq.~\eqref{eq.rotmatrix}, acting upon each of the constituent basis vectors:
\begin{equation} \label{eq.bdet}
	\boldsymbol{B}_\text{det} 
	= \mathcal{R}(\delta, \unitvector{s}_2) \mathcal{R}(\gamma, -\unitvector{s}_1) \mathcal{I}
	= \mathcal{R}(\delta, \unitvector{s}_2) \mathcal{R}(\gamma, -\unitvector{s}_1)
\end{equation}
In the laboratory frame, the matrix expressions for these two rotation operators are given by Eq.~\eqref{eq.rotmatrix} with $\unitvector{s}_1 = [ 1~0~0]^T$ and $\unitvector{s}_2 = [0~1~0]^T$:
\begin{equation}
  \mathcal{R}(\gamma, -\unitvector{s}_1) 
  = \left[
    \begin{matrix}
      1 & 0 & 0 \\
      0 & \cos\gamma & \sin\gamma \\
      0 & -\sin\gamma & \cos\gamma
    \end{matrix} 
  \right] 
  \qquad\text{and} \qquad
  \mathcal{R}(\delta, \unitvector{s}_2) = \left[
    \begin{matrix}
      \cos\delta & 0 & \sin\delta \\
      0 & 1 & 0 \\
      -\sin\delta & 0 & \cos\delta
    \end{matrix}
  \right]	 
  \nonumber
\end{equation}
leading to 
\begin{equation}
  \boldsymbol{B}_\text{det} = 
  \left[
    \begin{matrix}
      \cos\delta & -\sin\gamma\sin\delta & \cos\gamma\sin\delta \\
      0 & \cos\gamma & \sin\gamma \\
      -\sin\delta & -\cos\delta\sin\gamma & \cos\delta\cos\gamma
    \end{matrix}
  \right].	 
  \label{eq.deltagamma_expr}
\end{equation}
We note from Fig.~\ref{fig:coord_frame} that the negative sign in the $\gamma$-rotation above is necessary since the motor configuration at 34-ID-C results in a clockwise rotation about the positive $\unitvector{s}_1$-direction.
The columns of $\boldsymbol{B}_\text{det}$ in Eq.~\eqref{eq.deltagamma_expr} denote the unit-norm axes $\unitvector{k}_1$, $\unitvector{k}_2$, $\unitvector{k}_3$ of the detector frame, each expressed in the laboratory frame.
We note  that the first two columns of $\boldsymbol{B}_\text{det}$ are also the directions of Fourier space sampling vectors $\boldsymbol{q}_i$ and $\boldsymbol{q}_j$ from Fig.~\ref{fig:coord_frame}.

We next derive the expression for the third sampling vector $\boldsymbol{q}_k$.
The location of the Bragg peak $\boldsymbol{q}_0$ in Fourier space is computed analytically using $\boldsymbol{k}_i$ and $\boldsymbol{k}_f$ in the following manner:
\begin{align}
	\boldsymbol{k}_i &= \frac{1}{\lambda} \unitvector{s}_3 \tag{$\unitvector{s}_3$ is the downstream direction} \\
	\boldsymbol{k}_f &= \frac{1}{\lambda}\underbrace{\left[
		\mathcal{R}(\delta, \unitvector{s}_2) \mathcal{R}(\gamma, -\unitvector{s}_1)
	\right]}_{\text{composite rotation operator}} \unitvector{s}_3 \notag \\
	\Longrightarrow \boldsymbol{q}_0 &= \boldsymbol{k}_f - \boldsymbol{k}_i
		= \frac{1}{\lambda} \left[\mathcal{R}(\delta,\unitvector{s}_2) \mathcal{R}(\gamma, -\unitvector{s}_1) - \mathcal{I}\right] \unitvector{s}_3 \label{eq.Q0}
\end{align}
As mentioned earlier, the magnitude of $\boldsymbol{q}_k$ is given by the sweep step of the reciprocal lattice vector $\boldsymbol{q}_0$ due to the incremental rotation of the scatterer. The rotation in question is determined by the single angular step $\Delta\theta$ about  the $\unitvector{s}_2$ axis according to Fig.~\ref{fig:coord_frame} (we note that this is not always the case, for example in Ref.~\cite{Cha2016}, the object rotation is about the $\unitvector{s}_1$ axis). The change in $\boldsymbol{q}_0$ is  given by:
\begin{align}
  \Delta\boldsymbol{q}_0 
	&:= \mathcal{R}(\Delta\theta,
    \unitvector{s}_2)\boldsymbol{q}_0 -
    \boldsymbol{q}_0
	= \left[\mathcal{R}(\Delta\theta, \unitvector{s}_2) - \mathcal{I}\right] \boldsymbol{q}_0 \label{eq.DQ0} \\
	&= \frac{1}{\lambda}\left[
    \mathcal{R}(\Delta\theta, \unitvector{s}_2) - \mathcal{I}
    \right]
    \left[
    \mathcal{R}(\delta,\unitvector{s}_2) \mathcal{R}(\gamma, -\unitvector{s}_1) - \mathcal{I}
    \right] \unitvector{s}_3 \tag{using Eq.~\eqref{eq.Q0}}
\end{align}
We finally note that regardless of how $\boldsymbol{q}_0$ is rotated while rocking the scatterer, $\boldsymbol{q}_k$ is the displacement of the measurement plane relative to the center of the coherent intensity distribution, and therefore the \emph{negative} of $\Delta \boldsymbol{q}_0$. 
Keeping in mind that the norms of $\boldsymbol{q}_i$ and
$\boldsymbol{q}_j$ are both $p /\lambda D$ from 
Section~\ref{S:prelim}, we write down the simplified 
final expressions for the sampling vectors in 
Fourier space, \emph{still expressed in the laboratory frame}:
\begin{align}
	\boldsymbol{q}_i &:= \frac{p}{\lambda D}\unitvector{k}_1 = \frac{p}{\lambda D} \left[
		\begin{matrix}\cos\delta \\ 0 \\ -\sin\delta\end{matrix}
	\right] \label{eq.qi}\\
	\boldsymbol{q}_j &:= \frac{p}{\lambda D}\unitvector{k}_2 = \frac{p}{\lambda D} \left[
		\begin{matrix}-\sin\gamma\sin\delta \\ \cos\gamma \\ -\cos\delta\sin\gamma\end{matrix}
	\right] \label{eq.qj} \\
	\boldsymbol{q}_k &:= -\Delta\boldsymbol{q}_0 = -\frac{1}{\lambda} \left[
		\begin{matrix}
			\sin\delta\cos\gamma\left(\cos\Delta\theta - 1\right) + \sin\Delta\theta\left(\cos\delta\cos\gamma-1\right)\\
			0 \\
			\left(\cos\delta\cos\gamma-1\right)\left(\cos\Delta\theta-1\right) - \cos\gamma\sin\delta\sin\Delta\theta
		\end{matrix}
	\right] \label{eq.qk} \\
	&= \frac{\Delta\theta}{\lambda}
	\left[
		\begin{matrix}
			1 - \cos\gamma\cos\delta \\
			0 \\
			\cos\gamma\sin\delta
		\end{matrix}
	\right] + \mathcal{O}\left(\Delta\theta^2\right).
\label{eq.qk_fo}
\end{align}
Eq.~\eqref{eq.qk_fo} highlights the first-order dependence of $\boldsymbol{q}_k$ on the small rocking step $\Delta\theta$, obtained through a Taylor series expansion.
The sampling basis matrix $\boldsymbol{B}_{\text{recip}}$ is obtained by concatenating the numerically evaluated expressions for the sampling vectors: $\boldsymbol{B}_\text{recip} = \left[\boldsymbol{q}_i~\boldsymbol{q}_j~\boldsymbol{q}_k\right]$. 
The relations~\eqref{eq.qi},~\eqref{eq.qj} and~\eqref{eq.qk} explicitly demonstrate the highly intricate relationship between the  experimental considerations such as the scattering and sample  rotation geometries, and the manner in which Fourier space is discretely sampled. 
Specifically, in the Bragg geometry, the projections $\boldsymbol{q}_i^T \boldsymbol{q}_k$ and $\boldsymbol{q}_j^T \boldsymbol{q}_k$  cannot simultaneously be zero, implying that in BCDI, the sampling grid in Fourier space is inevitably non-orthogonal. 
The computation of the discrete Fourier-space points spanned by $\boldsymbol{B}_\text{recip}$ for a variety of standard goniometer geometries is in fact the primary function of the software package \emph{\textbf{xrayutilities}}~\cite{Kriegner2013}.

We now examine Eqs.\eqref{eq.qi} and~\eqref{eq.qk} in the pathological case of $\delta = 0$ but $\gamma \neq 0$, for which we show that it is impossible to acquire a 3D BCDI signal. 
Under these conditions, the incident and exit beams lie in the vertical $(\unitvector{s}_2, \unitvector{s}_3)$ plane and Eqs.~\eqref{eq.qi} and~\eqref{eq.qk} become
\begin{align}
	\boldsymbol{q}_i &= \frac{p}{\lambda D} \left[
			\begin{matrix}
				1 \\ 0 \\ 0
			\end{matrix}
		\right] \label{eq.qi_bad} \\
	\boldsymbol{q}_k &= -\frac{1}{\lambda} \left[
			\begin{matrix}
				\left(\cos\gamma - 1\right)\Delta\theta \\ 0 \\ 0
			\end{matrix}
		\right] + \mathcal{O}\left(\Delta\theta^2\right). 
\label{eq.qk_bad}
\end{align}
From Eqs.~\eqref{eq.qi_bad} and~\eqref{eq.qk_bad}, in the approximation of small rocking steps $\Delta\theta$ about the $\unitvector{s}_2$-direction, we deduce that two of the three Fourier-space sampling vectors are parallel and therefore not mutually linearly independent, rendering it impossible to sample a non-zero Fourier space volume for the 3D BCDI measurement.
Such a scenario more generally occurs when the rocking axis (in this case, $\unitvector{s}_2$) is improperly chosen to lie in the plane defined by $\boldsymbol{k}_i$ and $\boldsymbol{k}_f$.
For this reason, this axis is ideally chosen to lie well outside this plane in any BCDI measurement.
A particularly favorable case is when the rocking axis is perpendicular to this plane, a configuration sometimes referred to as a symmetric $\theta$-$2\theta$ geometry~\cite{Cha2016,Hruszkewycz2017a}.
In our special case of $\delta = 0$ and $\gamma \neq 0$, the symmetric $\theta$-$2\theta$ geometry dictates a rotational increment by the angular step $\Delta\theta$ about $\unitvector{s}_1$ instead of $\unitvector{s}_2$.
The rotation matrix $\mathcal{R}(\Delta\theta, \unitvector{s}_2)$ in Eq.~\eqref{eq.DQ0} is thus replaced with $\mathcal{R}(\Delta\theta, \unitvector{s}_1)$ in the analysis (a different sample rotation motor is typically chosen to achieve this in practice).
This results in the following modified expressions for the Fourier space sampling vectors:
\begin{align}
	\boldsymbol{q}_i^{(\theta\text{-}2\theta)} &= \frac{p}{\lambda D} \left[\begin{matrix}1 \\ 0 \\ 0\end{matrix}\right] \label{eq.qi_th2th} \\
		\boldsymbol{q}_j^{(\theta\text{-}2\theta)} &= \frac{p}{\lambda D} \left[\begin{matrix}0 \\ \cos\gamma \\ -\sin\gamma\end{matrix}\right] \label{eq.qj_th2th} \\
	\boldsymbol{q}_k^{(\theta\text{-}2\theta)} &= -\frac{1}{\lambda} 
	\left[
		\begin{matrix}
			0 \\
			\left(\cos\Delta\theta - 1\right)\sin\gamma - \left(\cos\gamma - 1\right)\sin\Delta\theta \\
			\left(\cos\gamma - 1\right)\left(\cos\Delta\theta - 1\right) + \sin\gamma\sin\Delta\theta
		\end{matrix} 
	\right] = \frac{\Delta\theta}{\lambda}
	\left[
		\begin{matrix}
			0 \\
			\cos\gamma - 1 \\
			-\sin\gamma
		\end{matrix}
	\right] + \mathcal{O}\left(\Delta\theta^2\right) \label{eq.qk_th2th}
\end{align}
We see from Eqs.~\eqref{eq.qi_th2th}, \eqref{eq.qj_th2th} and \eqref{eq.qk_th2th} that in the symmetric $\theta$-$2\theta$ geometry, the new sampling vectors are indeed non-coplanar, allowing one to query a finite 3D Fourier space volume.
This configuration is adopted in Bragg ptychography measurements and also in the main derivations of Part II.

In our derivations so far, we have chosen for visual clarity to express the experimental degrees of freedom and the eventual reconstruction in the universal frame $\boldsymbol{B}_\text{lab}$. 
In a completely equivalent treatment, the same analysis may also be developed entirely with respect to the detector frame $\boldsymbol{B}_\text{det}$ instead of $\boldsymbol{B}_\text{lab}$, provided the relevant vectors and rotation operators are formulated correctly.
This is in fact the natural frame of choice in Bragg ptychography applications and has been adopted in Part II, whose starting point is the theory developed so far.
In order to reconcile between these two frames we now provide a prescription to transform physical quantities seamlessly from one to the other.
 Any laboratory-frame vector $\boldsymbol{v}$ can be converted into the corresponding detector-frame representation $\boldsymbol{v}^\prime$ by projection along the axes of $\boldsymbol{B}_\text{det}$:
 \begin{equation}
\label{eq.to_det_frame}
 	\boldsymbol{v}^\prime = \boldsymbol{B}_\text{det}^T \boldsymbol{v}
 \end{equation}
with the reverse transformation from the detector- to the laboratory-frame representation also achieved in a straightforward manner:
\begin{align}
	\boldsymbol{v} &= \boldsymbol{B}_\text{det}^{-T} \boldsymbol{v}^\prime \notag \\
	&= \boldsymbol{B}_\text{det} \boldsymbol{v}^\prime
	\tag{since $\boldsymbol{B}_\text{det}$ is orthogonal}
\end{align}
Any rotation matrix $\mathcal{R}$ defined with respect to the laboratory frame may be transformed to its detector-frame representation $\mathcal{R}^\prime$ through the following similarity transformation:
\begin{equation}
	\mathcal{R}^\prime = \boldsymbol{B}_\text{det}^T \mathcal{R} \boldsymbol{B}_\text{det}.
	\label{eq.simtrans}
\end{equation}
It follows from Eq.~\eqref{eq.to_det_frame} that the laboratory-frame sampling basis $\boldsymbol{B}_\text{recip} = [\boldsymbol{q}_i~\boldsymbol{q}_j~\boldsymbol{q}_k]$ defined by Eqs.~\eqref{eq.qi},~\eqref{eq.qj} and~\eqref{eq.qk} is transformed to the detector frame by:
\begin{equation}
	\boldsymbol{B}_\text{recip} \xrightarrow[\text{det. frame}]{\text{to}} \boldsymbol{B}_\text{det}^T \boldsymbol{B}_\text{recip}
\end{equation}
where $\boldsymbol{B}_\text{det}$ is computed numerically from Eq.~\eqref{eq.deltagamma_expr}.
In either frame, the corresponding real-space sampling basis $\boldsymbol{B}_\text{real}$ of the final BCDI reconstruction may be computed from $\boldsymbol{B}_\text{recip}$ using Eq.~\eqref{eq.discretetransform}.
\vspace{.5em}

In Section~\ref{S:example} we describe an example of a BCDI reconstruction that implements the computational machinery that has been developed in this section. 
 
\section{An example: BCDI on an isolated nanoparticle}
\label{S:example}
With the the theoretical and computational machinery developed in Section~\ref{S:bcdi}, we are now in a position to demonstrate the effect of sampling-induced shear in the reconstruction of a real-world nano-particle imaged at a BCDI facility.
In our demonstrative example, the coherent diffraction from a compact, isolated nano-particle of silicon carbide (SiC) was collected at the 34-ID-C end station of the Advanced Photon Source.
This nano-particle was one of many nominally identical, tapered pillars with flat tops and bottoms, drop-cast on to a Si substrate after extraction from an etched SiC bulk single-crystal substrate.
A single such nano-particle was chosen for imaging purposes. 
The particulars of the experimental parameters during the BCDI measurement are given in Table~\ref{tab:experiment}.

\begin{figure}
	\centering
	\includegraphics[width=\hcolwidth]{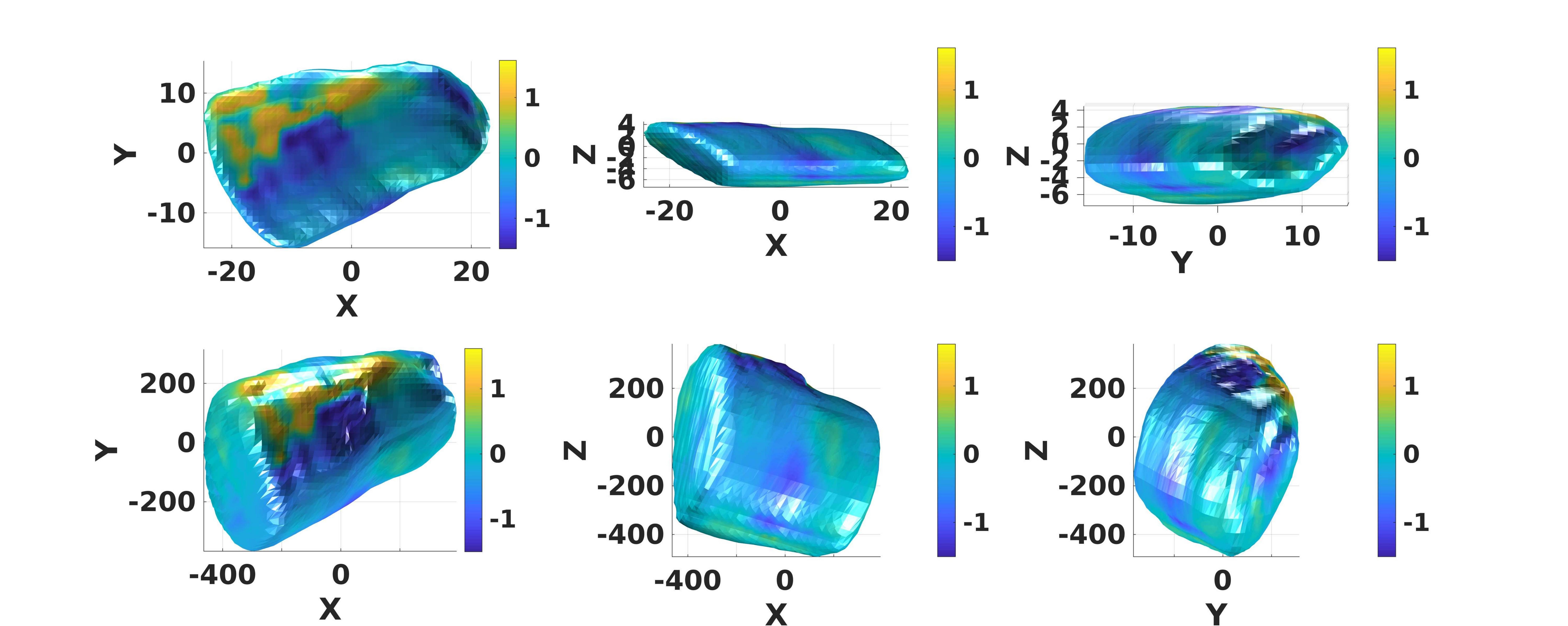}
	\caption{
		Isosurface plots of the reconstructed object ($XY$, $YZ$ and $XZ$ views), with the color scale depicting complex phase in radians.
		\textbf{Top row}: Direct isosurface plot of the scatterer from the phase retrieval solution array, without the required shear correction. Axis units are in pixels.
		\textbf{Bottom row}: Isosurface plots after the shear correction has been applied ($\boldsymbol{r} = \boldsymbol{B}_\text{real}\boldsymbol{m}$). Axis dimensions are in nanometers and the $X$, $Y$ and $Z$ axes corresponding to the laboratory-frame directions $\unitvector{s}_1$, $\unitvector{s}_2$ and $\unitvector{s}_3$ respectively.
	}
	\label{fig:views}
\end{figure}

\begin{figure}
	\centering
	\includegraphics[width=0.75\hcolwidth]{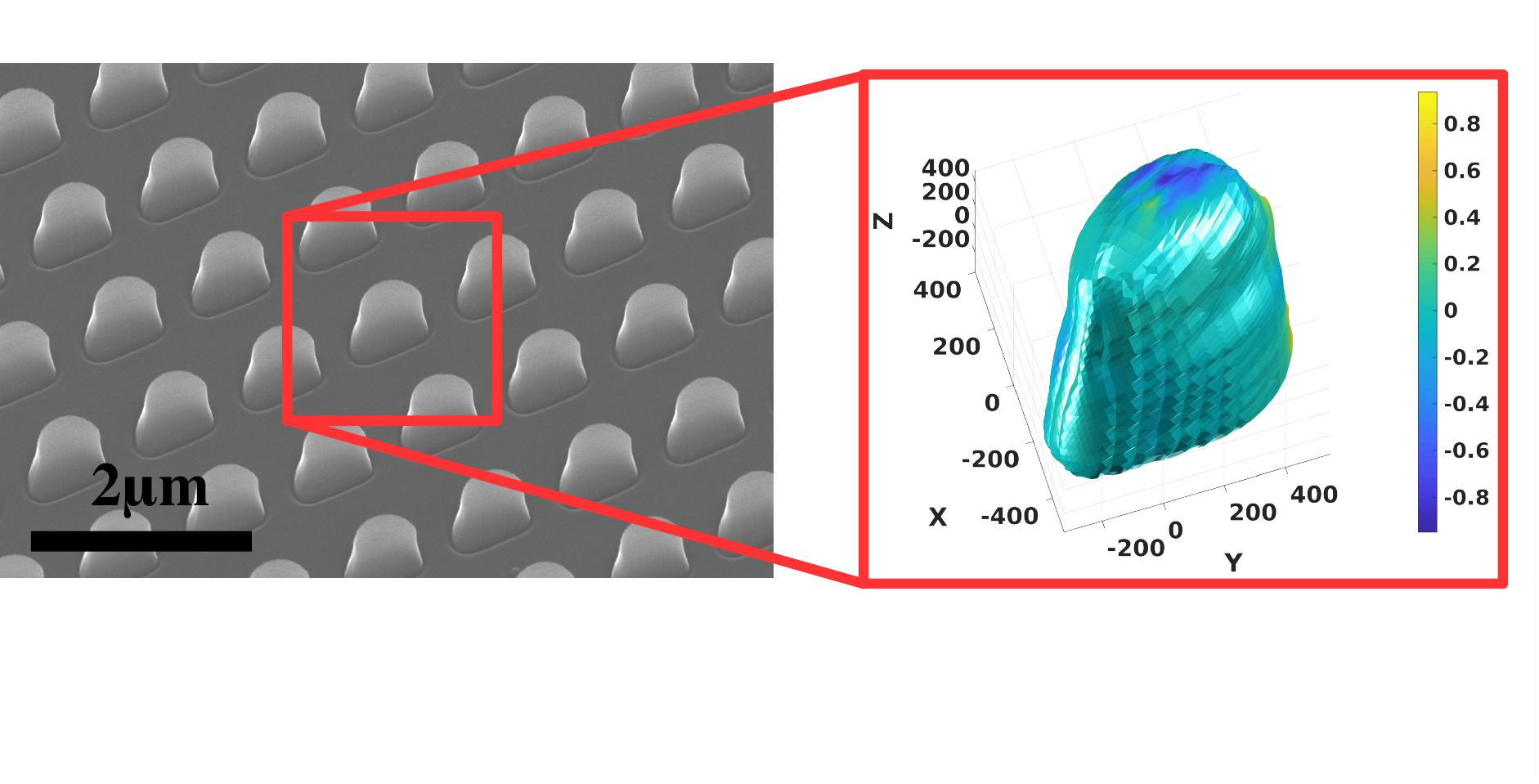}
	\caption{
		Comparison of the shear-corrected BCDI rendering of the SiC nano-particle with an SEM image. 
		In contrast to Fig.~\ref{fig:views}, this BCDI rendering has been artificially reoriented to match the view of the nano-particles in the SEM image, shown here on the original etched SiC block. 
		This view is no longer in the synchrotron laboratory frame.
		Here, the inherent mathematical degeneracy in the phase retrieval inverse problem (\emph{i.e.} if $\psi(\boldsymbol{r})$ is a real-space solution for an observed BCDI diffraction pattern, then so is $\psi^*(-\boldsymbol{r})$) was resolved by choosing the solution that most closely reproduced the asymmetric morphological features of the SiC particle in the SEM image, after application of the shear correction.
	}
	\label{fig:semcompare}
\end{figure}

Armed with this information, we may compute the following quantities in the laboratory frame: 
\begin{align}
	\boldsymbol{B}_\text{det} &= \left[
		\begin{matrix}
			0.869435 & -0.095149 & 0.484799 \\
			0 & 0.981279 & 0.19259 \\
			-0.494048 & -0.167445 & 0.853158
		\end{matrix}
	\right] \tag{from Eq.~\eqref{eq.deltagamma_expr}} \\
	\boldsymbol{B}_\text{recip} &
                                      = \left[
		\begin{matrix}
			173445.418 & -18981.475 & 42763.895 \\
			0 & 195757.552 & 0 \\
			-98558.742 & -33403.935 & 141174.943
		\end{matrix}
	\right] \text{ m}^{-1} \tag{from Eqs.~\eqref{eq.qi},
                                                  \eqref{eq.qj} and
                                                  \eqref{eq.qk}} 
\end{align}
and we finally have from Eq.~\eqref{eq.discretetransform}
\begin{equation}
  \nonumber
  \boldsymbol{B}_\text{real} = \boldsymbol{B}_\text{recip}^{-T} \left[
    \begin{matrix}
      256^{-1} && \\
      & 256^{-1} & \\
      && 100^{-1}
    \end{matrix}
  \right] = \left[
    \begin{matrix}
      19.214 & 0 & 34.340 \\
      0.870 & 19.955 & 13.642 \\
      -5.820 & 0 & 60.432
    \end{matrix}
  \right] \times 10^{-9} \text{ m }. 
\end{equation}
The columns of $\boldsymbol{B}_\text{real}$ above are the sampling
steps of the reconstructed scatterer corresponding to the pixels in the numerical reconstruction obtained from conventional phase retrieval. 
We further note that the real-space image thus rendered depicts the scatterer as it was oriented in the Bragg condition while in the diffractometer.
Fig.~\ref{fig:views} finally shows the effect of the shear correction on the rendered image of the scatterer.
A na\"ive isosurface rendering from the numerical array obtained from phase retrieval (top row) shows obvious distortions along different views of the nanocrystal image and the clear absence of top and bottom surfaces of the tapered pillar, as compared to the shear-corrected object (bottom row). 
Fig.~\ref{fig:semcompare} shows the shear-corrected view of the SiC nano-particle, reoriented to match an SEM image of the batch of SiC pillars prior to their release from the substrate. 
The essential morphological features in the SEM image are seen to be reproduced faithfully with the appropriate shear correction. 
In particular, the flat base of the pillar is clearly visible in the images in the bottom row.

\section{Summary}
\label{S:summary}
In Part I of this work we have described in general terms the scattering geometry of a BCDI experiment and its distortion effects on the imaged morphology of a crystalline scatterer obtained from phase retrieval.
This real-space distortion is demonstrated as an unavoidable effect of the non-orthogonal sampling of Fourier space using a conventional pixelated area detector and sample rocking arrangements.
We have provided a flexible numerical method to correct this image distortion, that can be easily implemented using standard linear algebra software packages and adapted to a variety of geometric configurations possible in BCDI. 
We have done this by examining the representations of real- and Fourier-space points and their fundamental conjugate relation through the Fourier transform. 

We have also demonstrated the validity of this shear correction with a BCDI reconstruction of a carefully fabricated silicon carbide nano-particle, corroborated with SEM images. 
This work serves as a theoretical basis for the analysis of BCDI diffraction geometry, as well as a general guideline for developing software tools for three-dimensional reconstruction.

The distortion correction formalism laid out in Part I unifies various customized prescriptions currently found in literature and in regular use at BCDI and ptychography facilities around the world.
As presented, it permits the flexible implementation of BCDI shear correction methodology to the experimental configurations of new BCDI beamlines, anticipating the wider adoption of BCDI at upcoming fourth-generation synchrotron light sources.
The formalism presented is the basic foundation of the methods developed in Part II, for direct reconstruction of the scatterer image on an orthogonal grid within the phase retrieval process.
This latter capability is demonstrated for the cases of even as well as uneven signal sampling in Fourier space, greatly increasing the scope of applicability of 3D phase retrieval.
An entirely new class of BCDI experiments potentially stand to benefit from this enhanced reconstruction capability, for instance measurements on dynamically varying samples or BCDI in the presence of unstable or vibrating components~\cite{Calvo-Almazan2019}.
As we shall see in Part II, such reconstructions can be achieved with minimal computational overhead through the modified 3D Fourier transform.

\appendix
\section{Computing strain components on a sheared grid}
\label{S:strain}
\begin{figure}
	\centering
	\includegraphics[width=\hcolwidth]{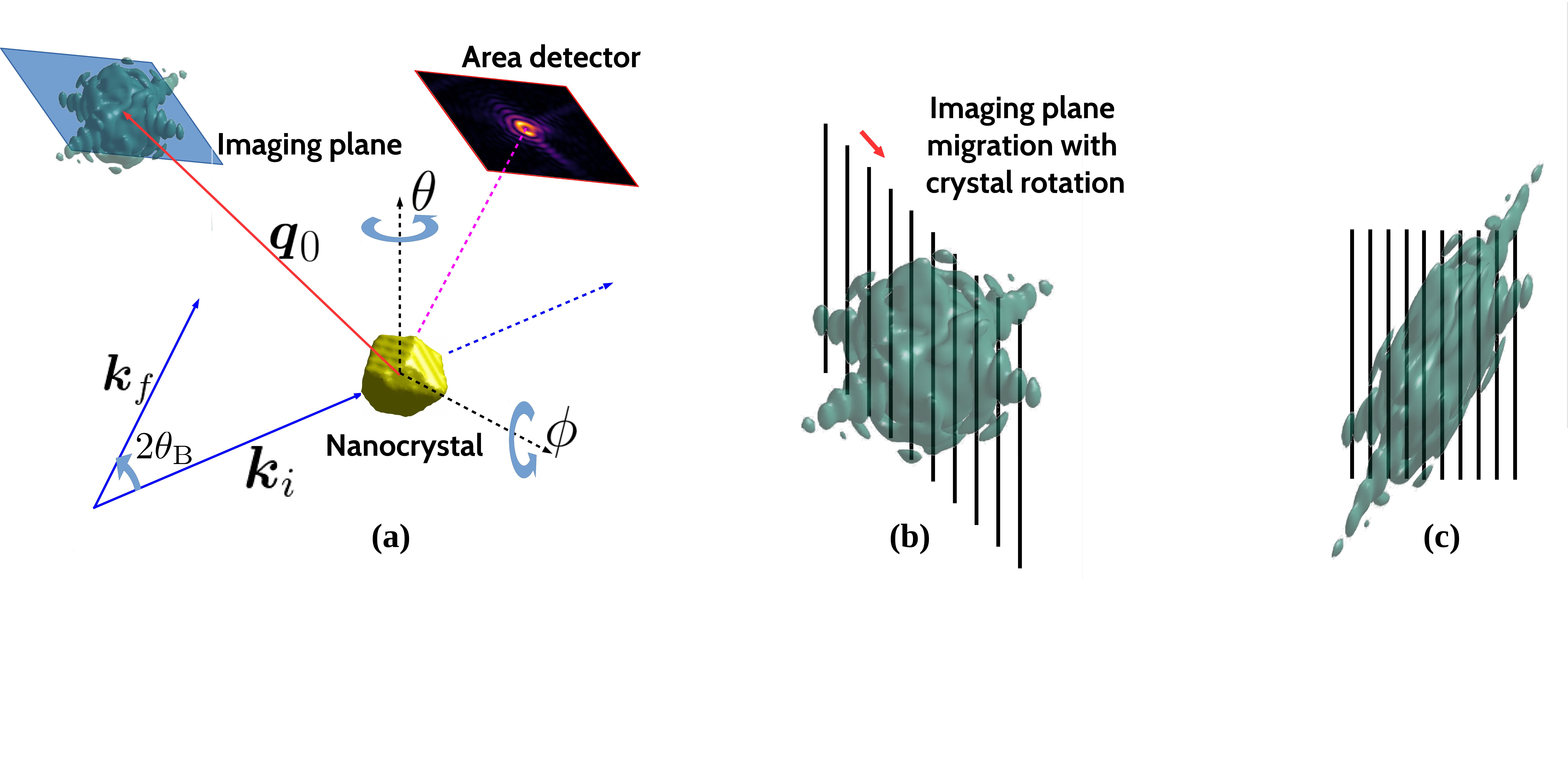}
	\caption{
		\textbf{(a)} Basic anatomy of a BCDI measurement of an isolated crystalline nano-particle.
		Rotating the scatterer in small increments (for instance about the $\theta$-direction) causes the reciprocal lattice point $\boldsymbol{q}_0$ of the scatterer's crystal structure to sweep an incremental angle in Fourier space.
		The `rocking' of the crystal's position about the Bragg condition effectively causes the measurement plane of the area detector to query parallel slices of the 3D coherent diffraction pattern. 
		\textbf{(b)} Effective shear in the relative Fourier space positions of the successive slices acquired by the area detector.
		The black lines indicate the position of the measurement plane relative to the center of the diffraction pattern.
		This sampling geometry is typical of crystal rocking about the $\phi$-axis in (a).
		\textbf{(c)} Inferred shape of the diffraction pattern if the collected detector images are na\"ively assumed to be orthogonal to each other.
		}
	\label{fig:concept}
\end{figure}

The components of the rank-2 strain tensor $\mathcal{E}$, when expressed in a convenient orthonormal frame, are typically indexed by two integers: $\epsilon_{ij}$.
Here the indices range over the number of dimensions ($i, j = 1, 2, 3$).
In BCDI, the component of the lattice strain field along the relevant reciprocal lattice vector $\boldsymbol{q}_0$ (see Fig.~\ref{fig:concept}(a)) at a point $\boldsymbol{r}$ in the crystal is given by:
\begin{equation}\label{eq.straintensor}
	\epsilon_{\boldsymbol{q}_0}(\boldsymbol{r}) 
	= \unitvector{q}_0^T \mathcal{E}(\boldsymbol{r}) \unitvector{q}_0
\end{equation}
where we have denoted $\unitvector{q}_0 \equiv [\hat{q}_{0,1}~~\hat{q}_{0,2}~~\hat{q}_{0,3}]^T$ as the unit-norm vector in the direction of $\boldsymbol{q}_0$ (\emph{i.e.} $\boldsymbol{q}_0 = \norm{\boldsymbol{q}_0} \unitvector{q}_0$).
If $\boldsymbol{u}(\boldsymbol{r})$ is the lattice distortion at the point $\boldsymbol{r}$, then Eq.~\eqref{eq.straintensor} can also be written as: 
\begin{align}
	\epsilon_{\boldsymbol{q}_0}(\boldsymbol{r}) &= \unitvector{q}_0^T \nabla\left[\unitvector{q}_0^T\boldsymbol{u}(\boldsymbol{r})\right] \label{eq.strainfromdisp} \\
	&= \left(\frac{1}{2\pi\norm{\boldsymbol{q}_0}}\right) \unitvector{q}_0^T \nabla \phi(\boldsymbol{r}) \label{eq.strainfromphase}
\end{align}
where $\nabla$ is the gradient with respect to $\boldsymbol{r}$ and $\phi(\boldsymbol{r}) \equiv 2\pi \boldsymbol{q}_0^T \boldsymbol{u}(\boldsymbol{r})$ is recognized as the complex phase field measured in a BCDI experiment.

We now wish to compute $\nabla \phi(\boldsymbol{r})$ at each grid point $\boldsymbol{r} = \boldsymbol{B}_\text{real} \boldsymbol{m}$, for use in Eq.~\eqref{eq.strainfromphase}. 
To do this, we first compute the projections of $\nabla \phi$ along the three independent sampling directions given by the columns of $\boldsymbol{B}_\text{real}$. 
If these directions are denoted by $\unitvector{e}_i^\prime$, where $i = 1, 2, 3$ and we define the integer array $\boldsymbol{m} \in \mathds{Z}^3$ as before, then these projections may be approximated by the finite differences of the discrete phase field, evaluated at $\boldsymbol{B}_\text{real} \boldsymbol{m}$:
\begin{align}
	\left(\unitvector{e}_1^\prime \right)^T \nabla \phi_{\boldsymbol{m}} &= 2\pi \left(
	\frac{
		\phi_{\boldsymbol{m}+[1~0~0]^T}-\phi_{\boldsymbol{m}}
	}
	{\delta x_1}
	\right)  + \mathcal{O}\left(\delta x_1\right) \label{eq.strainproj1}\\
	\left(\unitvector{e}_2^\prime \right)^T \nabla \phi_{\boldsymbol{m}} &= 2\pi \left(
	\frac{
		\phi_{\boldsymbol{m}+[0~1~0]^T}-\phi_{\boldsymbol{m}}
	}
	{\delta x_2}
	\right)  + \mathcal{O}\left(\delta x_2\right) \label{eq.strainproj2}\\
	\left(\unitvector{e}_3^\prime \right)^T \nabla \phi_{\boldsymbol{m}} &= 2\pi \left(
	\frac{
		\phi_{\boldsymbol{m}+[0~0~1]^T}-\phi_{\boldsymbol{m}}
	}
	{\delta x_3}
	\right)  + \mathcal{O}\left(\delta x_3\right) \label{eq.strainproj3}
\end{align}
where the $\delta x_i$ are the norms of the columns of $\boldsymbol{B}_\text{real}$, and $\phi_{\boldsymbol{m}}$ is shorthand for $\phi(\boldsymbol{B}_\text{real} \boldsymbol{m})$.
If we define $\boldsymbol{B}_r \equiv [\unitvector{e}_1^\prime~~\unitvector{e}_2^\prime~~\unitvector{e}_3^\prime]$ and the right-hand sides of Eqs.~\eqref{eq.strainproj1},~\eqref{eq.strainproj2} and~\eqref{eq.strainproj3} are concatenated to form a column vector $\boldsymbol{\xi}$, then we have $\boldsymbol{B}_r^T \nabla \phi(\boldsymbol{r}) = \boldsymbol{\xi}$ and therefore Eq.~\eqref{eq.strainfromphase} becomes:
\begin{equation}\label{eq.strainfinal}
	\boxed{
		\epsilon_{\boldsymbol{q}_0}(\boldsymbol{r}) = 
		\left(
			\frac{1}{2\pi\norm{\boldsymbol{q}_0}}
		\right) \unitvector{q}_0^T \boldsymbol{B}_r^{-T} \boldsymbol{\xi} =
		\left(
			\frac{1}{2\pi\norm{\boldsymbol{q}_0}}
		\right) \left(\boldsymbol{B}_r^{-1}\unitvector{q}_0\right)^T \boldsymbol{\xi}
	}
\end{equation}

Here $\boldsymbol{\xi}$ can be computed with relative ease owing to the availability of numerous software tools for finite differencing. 
Eq.~\eqref{eq.strainfinal} is a prescription to directly compute the strain component at each point $\boldsymbol{B}_\text{real}\boldsymbol{m}$ of the discrete, non-rectilinear grid spanned by $\boldsymbol{B}_\text{real}$ without having to interpolate the complex phase field on to a rectilinear grid in advance.

\ack{
The theoretical framework pertaining to the duality of continuous real- and Fourier-space was developed with support from the European Research Council (European Union's Horizon H2020 research and innovation program grant agreement No 724881).
Adaptation of this theory to discrete sampling and BCDI geometry and the accompanying x-ray measurements were supported by the U.S. Department of Energy, Office of Science, Basic Energy Sciences, Materials Science and Engineering Division. 
Sample preparation and SEM characterization made use of the Pritzker Nanofabrication Facility of the Institute for Molecular Engineering at the University of Chicago, which receives support from Soft and Hybrid Nanotechnology Experimental (SHyNE) Resource (NSF ECCS-1542205), a node of the National Science Foundation's National Nanotechnology Coordinated Infrastructure.
This research uses the resources of the Advanced Photon Source, a U.S. Department of Energy (DOE) Office of Science User Facility operated for the DOE Office of Science by Argonne National Laboratory under Contract No. DE-AC02-06CH11357.

 }

 \begin{table}
	\caption{Experimental parameters of the SiC nanocrystal BCDI scan, measured at Beamline 34-ID-C at the Advanced Photon Source. Refer to Fig.~\ref{fig:coord_frame} for the experimental geometry.}    
	\label{tab:experiment}
    \begin{tabular}{|c|c|c|}
    \hline
    \textbf{Parameter} & \textbf{Value} & \textbf{Description} \\ \hline
    $E$ & $9$ keV & Beam energy \\ \hline
    $\lambda$ & $1.378$ \AA & Wavelength \\ \hline
    $\Delta\theta$ & $0.0023^\circ$ & Angular increment \\ \hline
    $D$ & $2.0$ m & Object-detector distance \\ \hline
		$\gamma$ & $11.104^\circ$ & Detector alignment (elevation) \\ \hline
		$\delta$ & $29.607^\circ$ & Detector alignment (azimuth) \\ \hline
    $p$ & $55 \times 10^{-6}$ m & Pixel size \\ \hline
    $(N_1,N_2,N_3)$ & $(256,256,100)$ & Pixel array dimensions \\ \hline
    \end{tabular}
\end{table}
 
\end{document}